\documentclass{aa} 

\usepackage{graphicx}
\usepackage{txfonts}
\usepackage{multirow}
\usepackage{booktabs}
\usepackage{xcolor}

\begin{document}

   \title{A possibly inflated planet around the bright, young star DS\,Tuc~A
\thanks{Based on observations made with ESO Telescopes at the La Silla Paranal Observatory under programmes ID 
075.C-0202(A),   
0102.C-0618(A), 
0103.C-0759(A), 
075.A-9010(A), 	076.A-9006(A), 
073.C-0834(A), 
083.C-0150(B) 
} 
\fnmsep\thanks{This paper includes data collected by the TESS mission, which are publicly available from the Mikulski Archive for Space Telescopes (MAST).}}

\author{ S. Benatti,\inst{1,2}
D. Nardiello,\inst{3,1}
L. Malavolta,\inst{4}
S. Desidera,\inst{1}
L. Borsato,\inst{3,1} 
V. Nascimbeni,\inst{1,3}
M. Damasso,\inst{5}\\
V. D'Orazi,\inst{1,6} 
D. Mesa,\inst{1}
S. Messina,\inst{4} 
M. Esposito,\inst{7}
A. Bignamini,\inst{8}
R. Claudi,\inst{1} 
E. Covino,\inst{9} 
C. Lovis, \inst{10}
S. Sabotta\inst{7}
}

\institute{ INAF -- Osservatorio Astronomico di Padova, Vicolo dell'Osservatorio 5, I-35122, Padova, Italy  
\and INAF -- Osservatorio Astronomico di Palermo, Piazza del Parlamento, 1, I-90134, Palermo, Italy \\ 
    \email{serena.benatti@inaf.it}
\and Dipartimento di Fisica e Astronomia -- Universt\`a di Padova, Vicolo dell'Osservatorio 3, I-35122 Padova 
\and INAF -- Osservatorio Astrofisico di Catania, Via S. Sofia 78, I-95123, Catania, Italy 
\and INAF -- Osservatorio Astrofisico di Torino, Via Osservatorio 20, I-10025, Pino Torinese (TO), Italy 
\and Monash Centre for Astrophysics, School of Physics and Astronomy, Monash University, Clayton 3800, Melbourne, Australia
\and Th\"uringer Landessternwarte Tautenburg, Sternwarte 5, D-07778, Tautenburg, Germany  
\and INAF -- Osservatorio Astronomico di Trieste, via Tiepolo 11, I-34143 Trieste, Italy  
\and INAF -- Osservatorio Astronomico di Capodimonte, Salita Moiariello 16, I-80131, Napoli, Italy 
\and Observatoire de Gen\`eve, Universit\'e de Gen\`eve, 51 ch. des Maillettes, CH-1290 Sauverny, Switzerland
}

   \date{Received ; accepted }

 
  \abstract
{The origin of the observed diversity of planetary system architectures is one of the main topics of exoplanetary research. The detection of a statistically significant sample of planets around young stars allows us to study the early stages of planet formation and evolution, but only a handful of them is known so far. In this regard, a considerable contribution is expected from the NASA TESS satellite, which is now performing a survey of $\sim 85 \%$ of the sky to search for short-period transiting planets} 
{In its first month of operations, TESS found a planet candidate with an orbital period of 8.14 days around a member of the Tuc-Hor young association ($\sim$ 40 Myr), the G6V main component of the binary system DS\,Tuc. If confirmed, it would be the first transiting planet around a young star suitable for radial velocity and/or atmospheric characterization. We aim to validate the planetary nature of this companion and to measure its orbital and physical parameters. }
{We obtain accurate planet parameters by coupling an independent reprocessing of the TESS light curve with improved stellar parameters and the dilution caused by the binary companion; we analyse high precision archival radial velocities to impose an upper limit of about 0.1 M$_{\rm Jup}$ on the planet mass; we finally rule out the presence of external companions beyond 40 au with adaptive optics images. }
{We confirm the presence of a young, giant (${\rm R} = 0.50$ R$_{\rm Jup}$) planet having a not negligible possibility to be inflated (theoretical mass $\lesssim 20$ M$_{\oplus}$) around DS\,Tuc~A. We discuss the feasibility of mass determination, Rossiter-McLaughlin analysis and atmosphere characterization, allowed by the brightness of the star.
}
{}


\titlerunning{Young planet around DS\,Tuc~A}
\authorrunning{Benatti et al.}
   \maketitle
%

\section{Introduction}
\label{sec:intro}
After a few decades since the first detection, exoplanetary science is shifting from the discovery of exoplanets to understanding the origin of their astonishing diversity. The frequency of planets as a function of their mass, size, and host star properties, is a key parameter to test the diverse outcomes of planet formation and evolution. 
They are the result of the complex interplay between a variety of physical and dynamical processes operating on different timescales during formation and the successive orbital evolution. 
Those processes are in turn dependent on environmental conditions, such as disk properties, multiplicity, and stellar radiation fields, and involve mainly the migration processes. 
For instance, a smooth planet migration through the protoplanetary disk \citep{baruteau2014} can be revealed by small orbital eccentricities and spin-orbit alignments and operates in very short timescales (less than 10 Myrs, i.e., the disk lifetime), while the long term high-eccentricity migration (acting typically up to 1 Gyr, \citealt{chatterjee2008}) can result either in circular aligned orbits and short periods or eccentric misaligned orbits and long periods. Moreover, according to the so-called Coplanar High-Eccentricity Migration  \citep[CHEM, ][]{petrovich2015}, the existence of hot and warm Jupiters could be due to secular gravitational interactions with an eccentric outer planet in a nearly coplanar orbit, implying that a short period gaseous planet could be coupled to other distant bodies (of planetary or stellar nature).

In this context, the detection of young planets at short/intermediate orbital periods is crucial to investigate the regimes of planet migration. However, the high level of the stellar activity in young stars heavily hampers the detection of planetary signals with the radial velocity (RV) method, and the estimates on the frequency provided to-date \citep{donati2016,yu2017} still rely on small numbers and suffer for claimed detections that are not confirmed by independent investigations (e.g., \citealt{carleo2018}).

The NASA Transiting Exoplanet Survey Satellite (TESS; \citealt{2015JATIS...1a4003R}) is expected to play a crucial role in this framework. 
Being a space-borne full-sky survey, for the first time it will produce precise light curves for thousands of young stars, providing planetary candidates that however will require external validation.

In this paper, we validate the first young planet candidate spotted by TESS around one of the two components of the DS\,Tuc binary, the G6V star DS\,Tuc~A (HD\,222259, V=8.47), member of the Tuc-Hor association. 
This target (TESS Input Catalog ID: TIC 410214986) was observed in the first sector of TESS (2018 July 25 -- August 22), in both short (2 minutes) and long (30 minutes) cadence and has been tagged as TOI-200, i.e., TESS Objects of Interest (see the TOI releases webpage\footnote{\url{https://tess.mit.edu/toi-releases/}}), since transit signatures have been found.

In the following, we confirm the planetary nature of the potential candidate around DS\,Tuc~A by using data available in public archives. At first, we revise the stellar parameters of the host (Sect. \ref{sec:res}), then we propose our analysis of the TESS light curve, including a transit fit that considers the impact of the stellar companion on the estimation of the planet parameters (Sect. \ref{sec:phot}), and present our evaluation on RVs and adaptive optics data useful to constrain the planetary mass and the presence of additional bodies in the system (Sect. \ref{sec:val}). 
Discussion and conclusions are finally drawn in Sect. \ref{sec:conc}.


\section{Stellar parameters}
\label{sec:res}

DS\,Tuc is a physical binary formed by a G6 and a K3 component at a projected separation
of 5.4 arcsec (240 au at a distance of 44 pc).
The stellar parameters, some of which updated in our study, are listed in
Table \ref{t:star_param}.

The system is included among the core members of the Tuc-Hor association (e.g., \citealt{zuckerman2000}).
High levels of magnetic activity, a strong 6708\AA~~Lithium line, and its position 
on the color-magnitude diagram, slightly above the main sequence, strongly support a young age.
The kinematic analysis, based on Gaia DR2 results and the on-line BANYAN $\Sigma$ tool 
\citep{gagne2018}, yields a 99.9\% probability of membership in the Tuc-Hor association.
Its age is estimated to be 45$\pm$4 Myr from isochrone fitting \citep{bell2015}, $\sim$ 40 Myr from lithium depletion boundary \citep{kraus2014}, and $36^{+1.2}_{-1.3}$ from kinematics \citep{2019arXiv190207732C}.
We adopt in the following the 40 Myr as a mean value of these three determinations based on independent methods. 

\subsection{Spectroscopic analysis}
In order to get a metallicity estimate of the system, we have carried out spectroscopic analysis of DS\,Tuc~A, by exploiting a high-resolution, high signal-to-noise spectrum acquired with FEROS (nominal resolution R=48 000; \citealt{kaufer99}). The spectrum was reduced with a modified version of the ESO pipeline,
as described in \citet{desidera2015}.
Abundance analysis for very young stars (age $\lesssim$ 50 Myr) 
is not straightforward and presents several issues that hamper a standard approach. For instance, microturbulent velocity (v$_t$) values for this kind of stars are found to be extremely large, 
reaching 2.5 km~$s^{-1}$. This fact might be due to the presence of hot chromospheres, which affect the strong Fe~{\sc i} lines forming in the upper atmosphere layers (note that a trend of Fe abundances with the line formation depth has been found by e.g., \citealt{reddy17}). 
An extensive discussion of this topic will be presented in a dedicated series of papers, focussing on abundance determination of pre-main sequence stars (D'Orazi et al. 2019, in prep; Baratella et al. 2019, in prep.). Here we briefly describe our new approach:
effective temperature ($T_{\rm eff}$) has been fixed from photometry, whereas we have used Ti lines to optimise microturbulence and gravity by imposing the ionisation equilibrium condition such that logn(Ti~{\sc i}) and logn(Ti~{\sc ii}) abundances agree within the observational uncertainties. The exploitation of Ti lines present two main advantages: (i) the oscillator strengths of the lines ($\log gf$) have been experimentally determined by \cite{lawler13}, and (ii) the titanium lines form at an average formation depth of $<\log \tau_5> \sim -1$ (\citealt{gurtovenko15}), thus quite deep in the stellar photosphere. As a consequence, they are not strongly affected by chromospheric activity. 

Our results provide $T_{\rm eff} = 5542 \pm 70$ K, $\log g = 4.60 \pm 0.1 $ dex, $v_t = 1.18 \pm 0.10$ km~s$^{-1}$, logn(Ti\,{\sc i}) =$ 5.02 \pm 0.08$ (rms), logn(T\,{\sc ii}) =$ 84.94 \pm 0.07$. We inferred a metallicity [Fe/H]=$-0.08\pm0.02\pm0.06$, where the first uncertainty is the 
error on the mean from the Fe~{\sc i} lines, and 0.06 dex uncertainty is related to the atmospheric stellar parameters. The slightly sub-solar metallicity is consistent with a previous determination for Tuc-Hor association by \citet{vianaalmeida2009}.
We refer the reader to our previous works for details on error computation (e.g., \citealt{dorazi17}).
Unfortunately, we could not perform a spectroscopic parameter determination for the DS\,Tuc B component. This is because at the relatively low temperature ($T_{\rm eff} \sim 4600 K$), over-ionisation effects for iron and titanium lines are at work in stars younger than roughly 100 Myr. The difference between
logn(Fe~{\sc i}) and logn(Fe~{\sc ii}) and/or logn(Ti~{\sc i})and logn(Ti~{\sc ii}) are larger than ~0.8 dex, resulting in not reliable (unphysical) parameters and abundances.
This kind of over-ionisation and over-excitation effects have been previously noticed in several young clusters, such as e.g., NGC 2264 (\citealt{king00}), the Pleaides (\citealt{king00}; \citealt{schuler10}), M34 (\citealt{schuler03}), IC 2602 and IC 2391 (\citealt{dorazi09}).

\subsection{Rotation period} \label{a:rot}
\begin{figure}[htbp]
\centering
\includegraphics[width=0.65\columnwidth,angle=90]{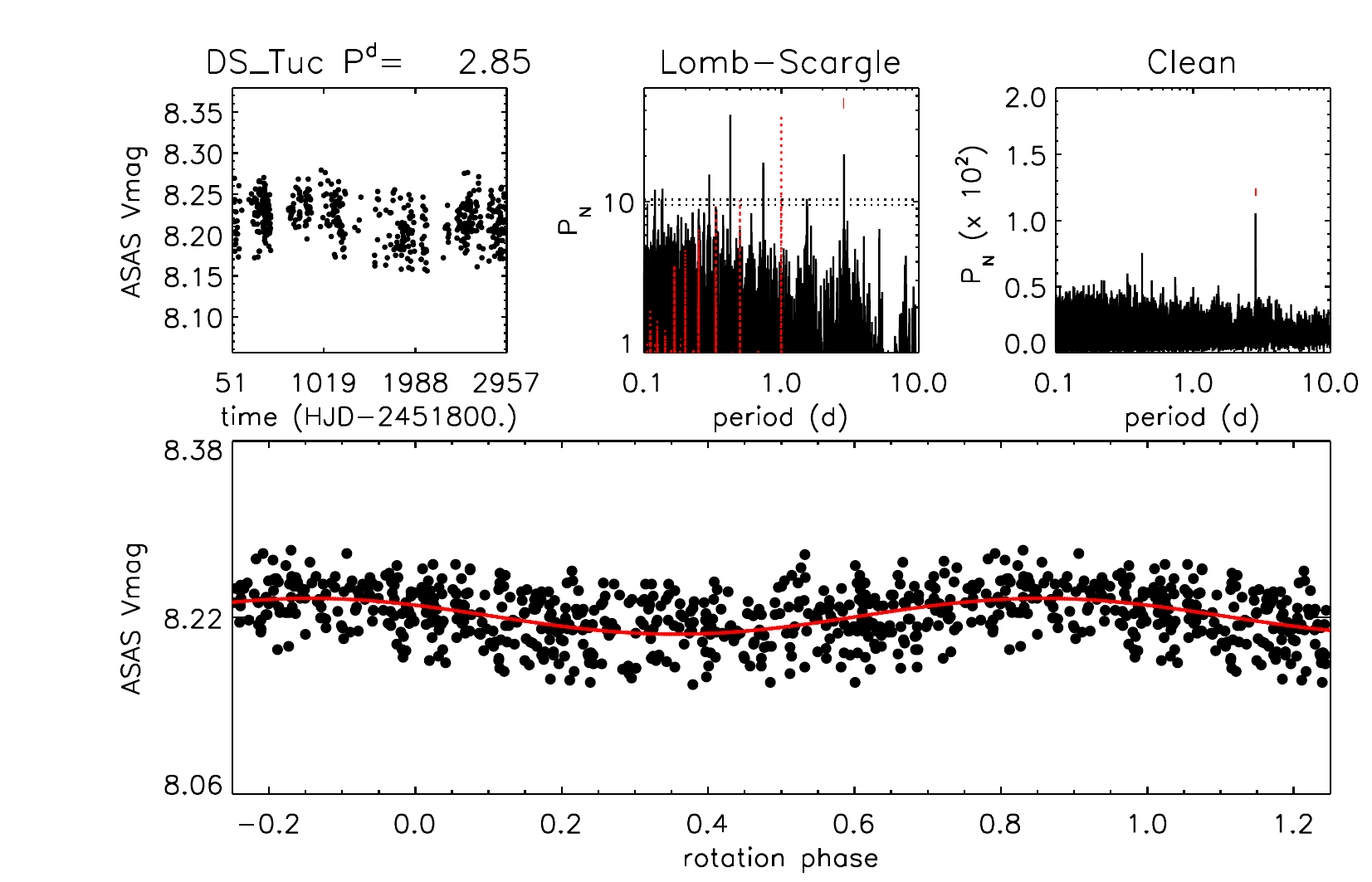}
\caption{\label{fig:a:SMe} Time series, periodogram, and phased curve of the ASAS photometric data of DS\,Tuc, collected between 2000 and 2008.}
\end{figure}

DS\,Tuc was observed as part of the ASAS (All Sky Automated Survey; \citealt{1997AcA....47..467P}) survey from 2000 to 2008. From the public archive, we retrieved a total of 560 V-band measurements with an average photometric accuracy $\sigma = 0.031$ mag. The time series is shown in the upper left panel in Fig. \ref{fig:a:SMe}, where a long term photometric variation, probably due to a spot cycle, is clearly seen. The complete time series was analysed with the Lomb-Scargle \citep{1982ApJ...263..835S} and CLEAN \citep{Roberts87} periodograms (Fig. \ref{fig:a:SMe}, upper central and right panels). Both methods returned a rotation period P = 2.85$\pm$0.02 days with a False Alarm Probability (FAP) $<$ 0.01. The FAP was computed with Monte Carlo simulations (e.g. \citealt{messina2010}), whereas the uncertainty on the rotation period was derived following \cite{Lamm04}. Periodograms were also computed for segments of the whole time series and the same rotation period was retrieved in five out of 12 segments. 

An independent measure of the $P_\mathrm{rot}$ value has been obtained from the TESS short cadence light curve extracted by us (see the details in Sect. \ref{app:lc}), after masking the transit events. The autocorrelation function analysis (e.g., \citealt{McQuillan2013}) shows the first peak at 2.91 days, while the Generalised Lomb-Scargle (GLS) periodogram \citep{2009A&A...496..577Z} recovered the expected signal at 2.85 days with an amplitude of 16 millimags in the TESS passband, after removing the contribution of the binary companion. Finally, we measured the rotational period with its error by modelling the light curve using a Gaussian process (GP) with a quasi-periodic kernel. We made use of the {\tt PyORBIT} code \citep{Malavolta2016,Malavolta2018}, following the prescriptions in \cite{Rice2019}, with the GP computed by the {\tt george} package \citep{Ambikasaran2015}. Following \cite{LopezMorales2016}, we imposed a prior on the coherence scale to allow a maximum of three peaks per rotational period\footnote{The proposed value of $0.50 \pm 0.05$ converts to $0.35 \pm 0.03$ due to the different mathematical formulation used in {\tt PyORBIT}.}. To speed up the computation, we binned the light curve in step of 1 hour. We measured a rotation period of $P_\mathrm{rot} = 2.99 \pm 0.03$ days, with a covariance amplitude of $\lambda_\mathrm{TESS} = 11 \pm 2$ millimag and an active region decay time scale of $P_\mathrm{dec} = 3.12 \pm 0.20$ days. The fact that the GP and the autocorrelation function are more sensitive to the evolution of the active regions explains for the slight discrepancy (of the order of 5\%) of the estimated P$_{\rm rot}$ with respect to GLS and LS periodograms methods.

In the following, we adopt as a rotation period of the target the value obtained from ASAS data.

\subsection{Parameters} \label{sec:param}
The radii and masses of the components were derived using the PARAM interface \citep{dasilva2006}.
By fixing conservative age boundaries between 30 to 60 Myr from the
Tuc-Hor membership, adopting the effective temperature (with an error of 70 K) from \citet{pecaut2013}, and the spectroscopic value of metallicity, we obtain mass and radius of 0.96 M$_{\odot}$ and 0.87 R$_{\odot}$ for DS\,Tuc~A, and 0.78 M$_{\odot}$ and 0.77 R$_{\odot}$, respectively, for the B component.
Our stellar radius is very similar to the one derived in Gaia DR2.

The stellar radius of the primary, coupled with the observed rotation period yields 
a rotational velocity of 15.5 km s$^{-1}$, in agreement with our determination of $15.5 \pm 1.5$ km s$^{-1}$ (from spectral synthesis, as in e.g., \citealt{dorazi17}). This suggests that the star is seen close to equator-on.
The observed motion of the binary components on the plane of the sky 
is nearly radial, suggesting
a very eccentric and/or a very inclined (close to edge-on) orbit.

Finally, the star has no significant IR excess \citep{zuckerman2011}, which excludes prominent dust belts around it.

\begin{table}[htbp]
   \caption[]{Stellar properties of the components of DS\,Tuc.}
     \label{t:star_param}
     \small
     \centering
       \begin{tabular}{lccc}
         \hline
         \noalign{\smallskip}
         Parameter   &  \object{DS\,Tuc~A} & \object{DS\,Tuc~B}  & Ref. \\
         \noalign{\smallskip}
         \hline
         \noalign{\smallskip}
$\alpha$ (2000)          &  23:39:39.481 &  23:39:39.270 &  1   \\
$\delta$ (2000)          & -69:11:44.709 & -69:11:39.495 &  1   \\
$\mu_{\alpha}$ (mas/yr)  &  $79.464\pm0.074$  &  $78.022 \pm0.064$ & 1 \\
$\mu_{\delta}$ (mas/yr)  & $-67.440\pm 0.045$  & $-65.746 \pm0.037$ & 1 \\
RV     (km/s)            &   8.0$\pm$ 0.8  &   6.1$\pm$ 0.1 & 2 \\ 
RV     (km/s)            &   7.8815 $\pm$0.0081$^{\mathrm{a}}$   &  --  & 3 \\  
RV     (km/s)            &   $7.20\pm0.32$   &   $5.32\pm0.65$ & 1 \\
RV     (km/s)            &   7.4$\pm$0.2     &      --        & 4 \\
$\pi$  (mas)             & $22.6663\pm0.0354$ & $22.6504\pm0.0297$  & 1   \\
\noalign{\medskip}
V (mag)                       &   $8.469\pm 0.013$ &    $9.84\pm 0.01 $ & 5 \\
B-V (mag)                     &    $0.693\pm 0.017$   &   $1.00\pm 0.01 $  & 5 \\
V-I (mag)                &    $0.77 \pm 0.01$   &  $ 1.16\pm 0.01 $  & 2 \\
G (mag)                       & $ 8.3193\pm0.0010$  &  $9.3993\pm0.0014$ & 1 \\
BP-RP (mag)                   &  0.8908             &  1.2769   & 1 \\
J$_{\rm 2MASS}$ (mag)             &  $7.122\pm0.024$       &  $7.630\pm0.058$  & 6 \\
H$_{\rm 2MASS}$ (mag)             &  $6.759\pm0.023$       &  $7.193\pm0.034$  & 6 \\
K$_{\rm 2MASS}$ (mag)             &  $6.676\pm0.034$       &  $7.032\pm0.063$  & 6 \\
T (mag)                 &  $7.771\pm0.018$        &       $8.607 \pm 0.030$     & 7 \\
\noalign{\medskip}
Spectral Type                       &  G6V     &   K3Ve    &  2 \\
T$_{\rm eff}$ (K)            &  $5542\pm21$     &   $4653\pm18$    & 5  \\ 
T$_{\rm eff}$ (K)            &  $5597.5^{+28.0}_{-59.2} $ &  $4778.8^{+84.2}_{-61.0} $  &  1 \\ 
$\log g$                 &  ~~$4.60\pm0.15$   &    --     &  3 \\ 
${\rm [Fe/H]}$ (dex)          &  $-0.08\pm0.06$ &    --   &  3 \\ 
\noalign{\medskip}
$\log R^{'}_{\rm HK}$        &    $-4.166\pm0.015$  &    --     &  3 \\  
$\log R^{'}_{\rm HK}$        &    $-4.09$  &     --    &  8 \\ 
$ v \sin i $ (km/s)      &    $18.3 \pm 1.8    $      &  $ 14.6 \pm 1.5    $   &  2 \\  
$ v \sin i $ (km/s)      &    $15.5 \pm 1.5    $      &  --      &  3 \\  
${\rm P_{\rm rot}}$ (d)  &   2.85$\pm$0.02  &      --     & 3  \\
$\log L_{\rm X}$             &  \multicolumn{2}{c}{30.21$^{\mathrm{b}}$} & 3 \\
$\log L_{\rm X}/L_{\rm bol}$     &  \multicolumn{2}{c}{-3.35$^{\mathrm{b}}$} & 3 \\
EW Li (m\AA)                   &     216           &   232       & 2 \\
\noalign{\medskip}
Mass (M$_{\odot}$) &    0.959$\pm$0.031$^{\mathrm{c}}$  &  0.782$\pm$0.022$^{\mathrm{c}}$  &  3 \\
Radius (R$_{\odot}$) &  0.872$\pm$0.027$^{\mathrm{c}}$  &  0.769$\pm$0.033$^{\mathrm{c}}$  &  3 \\
Radius (R$_{\odot}$) &    $0.88^{+0.2}_{-0.1} $ &  $0.79^{+0.02}_{-0.03}$  &  1 \\
Age  (Myr)               &\multicolumn{2}{c}{$40\pm5$}  &  9  \\

         \noalign{\smallskip}
         \hline
      \end{tabular}

References: 1 \citet{2018A&A...616A...1G}; 
            2 \citet{torres2006};  
            3 This paper;
            4 \citet{nordstrom2004}; %
            5 \citet{pecaut2013}; 
            6 \citet{2006AJ....131.1163S};
            7 \citet{2018AJ....156..102S};
            8 \citet{henry1996};
            9 \citet{kraus2014}.
\begin{list}{}{}
\item[$^{\mathrm{a}}$] Offline HARPS DRS (see Sect. \ref{sec:rv})
\item[$^{\mathrm{b}}$] A+B 
\item[$^{\mathrm{c}}$] Derived using the PARAM web interface \citep{dasilva2006}
\end{list}
\end{table}

\section{TESS light curve analysis\label{sec:phot}}
In this Section, we present our extraction and analysis of the TESS light curve of DS\,Tuc.

\subsection{Light curve extraction} \label{app:lc}
\begin{figure*}[htbp]
\centering
  \includegraphics[width=1.5\columnwidth]{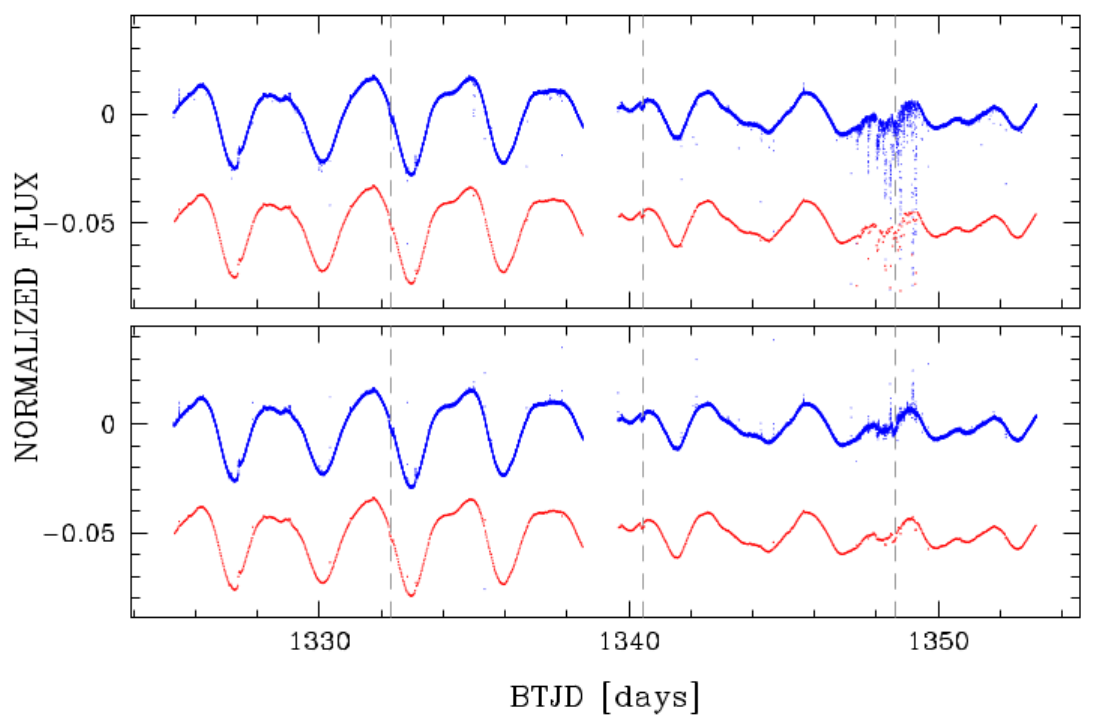}
  \caption{Light curve of DS\,Tuc before (top-panel) and after
    (bottom-panel) the correction of the systematic effects. Each
    panel shows the 2-minute (red points) and 30-minute (blue points)
    cadence light curve. BTJD is the Barycentric TESS Julian Day, corresponding to BJD - 2,457,000. The dashed lines represent the location of the three detected transits of DS\,Tuc~A~b. \label{fig:cotrend}}
\end{figure*}
DS\,Tuc was observed in the first sector of TESS for 27 days, falling on the CCD\,2 of Camera\,3.
We extracted the light curve from the 30-minute cadence full frame
images (FFIs) exploiting the TESS version of the software
\texttt{img2lc} developed by \cite{2015MNRAS.447.3536N} for ground-based
observations, and also used for Kepler images by \cite{2016MNRAS.456.1137L}. 
Briefly, using the FFIs, empirical Point Spread Functions
(PSFs), and a suitable input catalog (in this case Gaia~DR2, \citealt{2018A&A...616A...1G}), the routine performs aperture and
PSF-fitting photometry of each target star in the input catalog, after
subtracting the neighbors located between 1\,px and 25\,px from the
target star. A detailed description of the pipeline will be given by
Nardiello et al.~(in preparation).
As for 2-minute cadence light curve, we adopted the raw one
(\texttt{PDCSAP\_FLUX}) released by the TESS team and available for
download on the Mikulski Archive for Space Telescopes (MAST) archive.

We corrected both the 2- and 30-minute cadence light curves of DS\,Tuc
adopting the same procedure described by \cite{2016MNRAS.463.1831N}.  For
each star with $G_{\rm BP}<13$ located on the CCD\,2 of the Camera 3,
we computed the mean and the rms of its light curve. We selected the
25\,\% of the stars with the lowest rms, and we used them to extract a
Cotrending Basis Vector (CBV) adopting the Principal Component
Analysis. We applied the Levenberg–Marquardt method to find the
coefficient $A$ that minimises the expression $F_{\rm raw}^j-A \cdot
{\rm CBV}^j$, where $F_{\rm raw}^j$ if the raw flux of the star at
time $j$, with $j=1,...,N_{\rm epochs}$, and $N_{\rm epochs}$ the
number of points in the light curve. We cotrended the light curve
subtracting $A\cdot {\rm CBV}$ to the light curve of DS\,Tuc.
In Fig.~\ref{fig:cotrend} we show the 2-minute (blue points) and the
30-minute (red points) cadence before (top-panel) and after
(bottom-panel) the correction.
Three transit events can be found in the time series (dashed lines in Fig. \ref{fig:cotrend}), but the third one occurred during a temporary failure of the satellite pointing system ($1347 \leq {\rm TJD} \leq 1349$, where TJD = JD-2,457,000), with a consequent degradation of the data (upper panel of Fig. \ref{fig:cotrend}).
Thanks to our alternative detrending algorithm we were able to recover the third transit, although at lower quality with respect to the others. 

\begin{figure}[htbp]
\centering
  \includegraphics[width=\columnwidth]{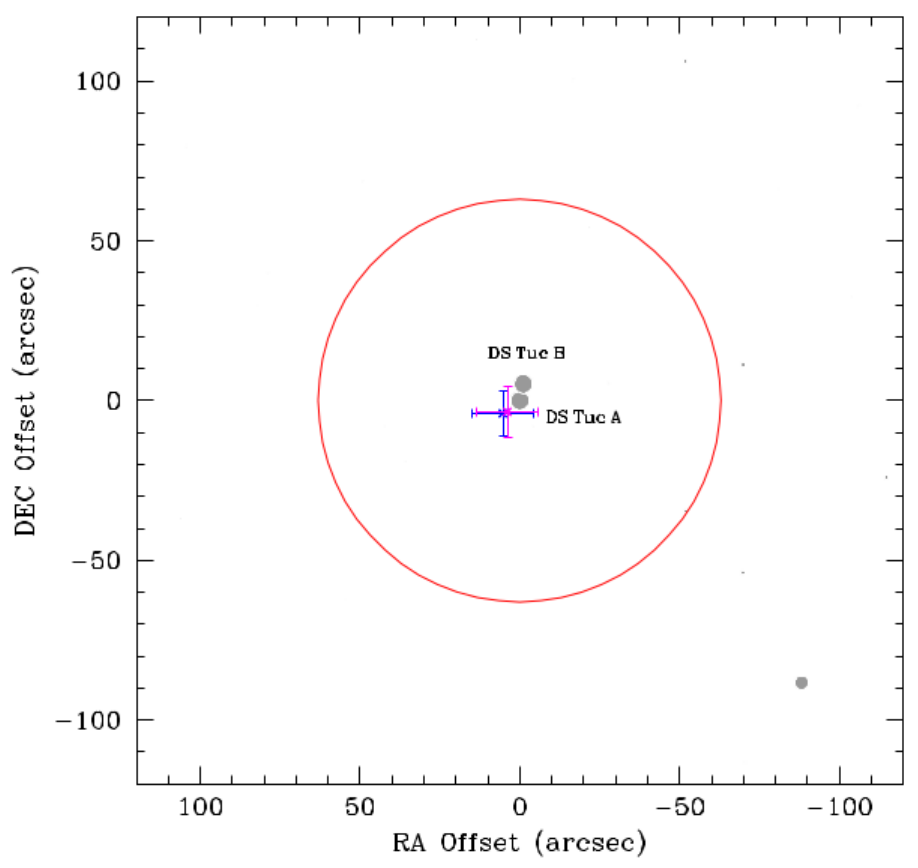}
  \caption{Difference image centroid offset for DS\,Tuc~A from TESS observations. Grey dots mark the position of the sources in the field, the red circle indicates the extension of the photometric aperture, while the magenta and blue crosses with error bars represent the offset with respect to the TESS out-of-transit images and the coordinates of the target from Gaia DR2.} \label{fig:centroid}
\end{figure}

The TESS light curve of DS\,Tuc shows evidence for transits of two distinct companions.
According to the Data Validation Report associated to each candidate (see \citealt{2018PASP..130f4502T}), one of the signals could be related to a companion with an orbital period of 20.88 days, but after a first inspection of the light curve, the results of the fitting procedure, and the matching of the stellar field with Gaia DR2 data (see also Sect. \ref{sec:imag}), it shows a very high probability of being a false positive, possibly due to an instrumental effect. The second one appears to be produced by a real companion, characterised by an orbital period of 8.138 days (semi-major axis = 0.09 au) and a planet radius of 8.3 R$_{\oplus}$, evaluated adopting the stellar radius from the TESS Input Catalog ($1.19 \pm 0.13$ ${\rm R_{\odot}}$, \citealt{2018AJ....156..102S}) which is significantly larger with respect to the one we obtained in Sect. \ref{sec:param} and also reported by Gaia. Moreover, since the projected separation of the two stellar companions of the binary system is 5.36", and the TESS pixel scale is 21"/pix, the two sources are not resolved, leading to a likely overestimation of the planetary radius as a consequence of the dilution of the transit by the binary companion.
Before proceeding, we verified that the transit of the 8-days period candidate occurred on the primary star by performing a centroid analysis from the TESS full-frame images. Fig. \ref{fig:centroid} depicts the considered field, limited by the 3-pixel aperture used for the flux extraction (red circle), with grey dots indicating the positions of DS\,Tuc~A and B and a faint background source. The plot shows the calculated offsets (and the corresponding error bars) with respect to the TESS out-of-transit images (magenta) and to the position reported in Gaia DR2 (blue).
In both cases, the offset falls within 1-sigma from the position of DS\,Tuc~A. 

In the next two Sections, we modelled the extracted light curve with two different approaches: the first one considered only the first two transits and simultaneously fits the rotational modulation. The second one considered all the available transits and a different method to flatten the curve. The two methods led to consistent results, but we decided to adopt the planet parameters from the first one. 

\subsection{Two transits fit} \label{sec:fit}
\begin{figure*}[htbp]
\centering
  \includegraphics[width=1.5\columnwidth]{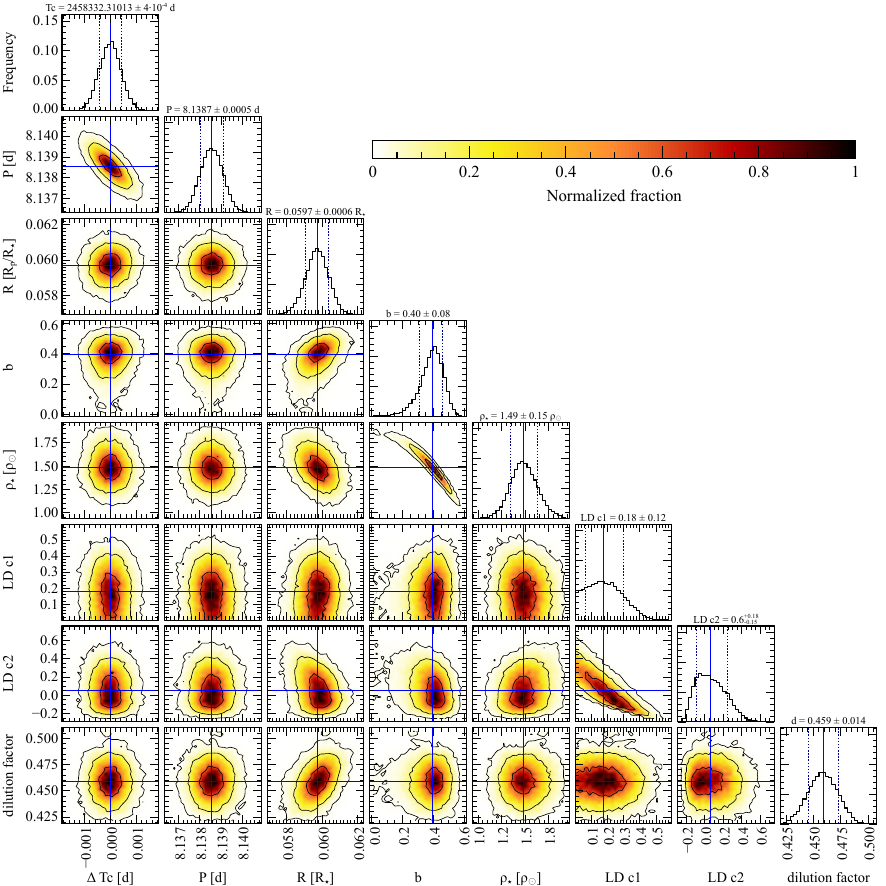}
  \caption{Posterior distribution of the 2 transits fit described in Sect. \ref{sec:fit}. \label{fig:corner}}
\end{figure*}

In the first approach, a temporal window of 0.6 days (roughly 5 times the duration of the transit) was selected around each mid-transit time, which allowed to approximate the rotational modulation with a second order polynomial for each transit. The analysis was performed with {\tt PyORBIT}, a convenient wrapper for the transit modelling code {\tt batman} \citep{Kreidberg2015} and the affine invariant Markov chain Monte Carlo sampler {\tt emcee} \citep{Foreman-Mackey2013} in combination with the global optimization code {\tt PyDE}\footnote{Available here: \url{https://github.com/hpparvi/PyDE}}. Our model included as parameters: the time of first transit $T_c$, the planetary period $P$, the impact parameter $b$, the planetary to stellar radius $R_\mathrm{p}/R_{\star}$ and the stellar density in unit of solar density $\rho_{\star}$. The dilution factor $F_\mathrm{B}/F_\mathrm{A}$ was included as free parameter to account for its impact on the error estimate of $R_\mathrm{p}$, with a Gaussian prior derived from the Gaia and J magnitudes of the A and B components of the stellar system transformed into the TESS system by using the relation in \cite{2018AJ....156..102S}\footnote{At the time of analysis and writing of this paper the new calibrations by \cite{2019arXiv190510694S} were not available.}. For each transit we also included a jitter term, quadratically added to the photon-noise estimate of TESS light curve, to take into account short-term stellar activity noise and unaccounted TESS systematics. We tested eight different models to fit the TESS light curves: without priors on the limb darkening (LD) coefficients $u_1$ and  $u_2$, prior only on $u_1$, prior only on $u_2$, and priors on both coefficient, with each model repeated twice for the circular and Keplerian orbit respectively. The priors on the LD coefficients were computed through a bilinear interpolation of the grid presented by \cite{2018A&A...618A..20C}, considering $T_{\rm eff}$, $\log g$ and metallicity in Table \ref{t:star_param}.
We followed the prescriptions of \cite{Kipping2013} for the parametrization of the LD coefficients, and the ($\sqrt{e} \sin \omega$, $\sqrt{e} \cos \omega$) parametrization for eccentricity and argument of pericenter \citep{Eastman2013}. We accounted for the 120 s exposure time of TESS when modelling the transit \citep{Kipping2010}. All the parameters were explored in linear space, with the exception of the photometric jitter terms. Our models have 16 free parameters for the circular orbit cases, and 18 for the keplerian ones. We deployed a number of walkers equal to eight times the dimensionality of the problems, for a total of {128} and {144} walkers. For each model we run the sampler for 50000 steps, we then applied a burn-in cut of 15000 steps and a thinning factor of 100, obtaining a total of 44800 and 50400 independent samples for the circular and keplerian models respectively. Confidence intervals of the posteriors were computed by taking the 34.135th percentile from the median (Fig. \ref{fig:corner}).
The fit was performed after rescaling the flux for its median value.
In addition to the MCMC fit, we performed the model selection by computing the Bayesian evidence with the nested sampling algorithm {\tt PolyChord} \citep{Handley2015}. Models with a prior on one or both the LD coefficients are disfavoured, although not significantly, with respect to the model without priors. We found that all the parameters are consistent across the models, with the exception of the LD coefficients that show a mismatch with the selected priors, favouring the choice of uninformative priors to exclude unphysical solutions (in agreement with \citealt{Kipping2013}). 
The discrepancy between the literature and derived values for the limb darkening coefficients is likely the consequence of the strong magnetic field of the star, as investigated by \citet{2015A&A...581A..43B} and confirmed in \citet{2018A&A...616A..39M}.

\begin{figure}[htbp]
  \includegraphics[width=\columnwidth]{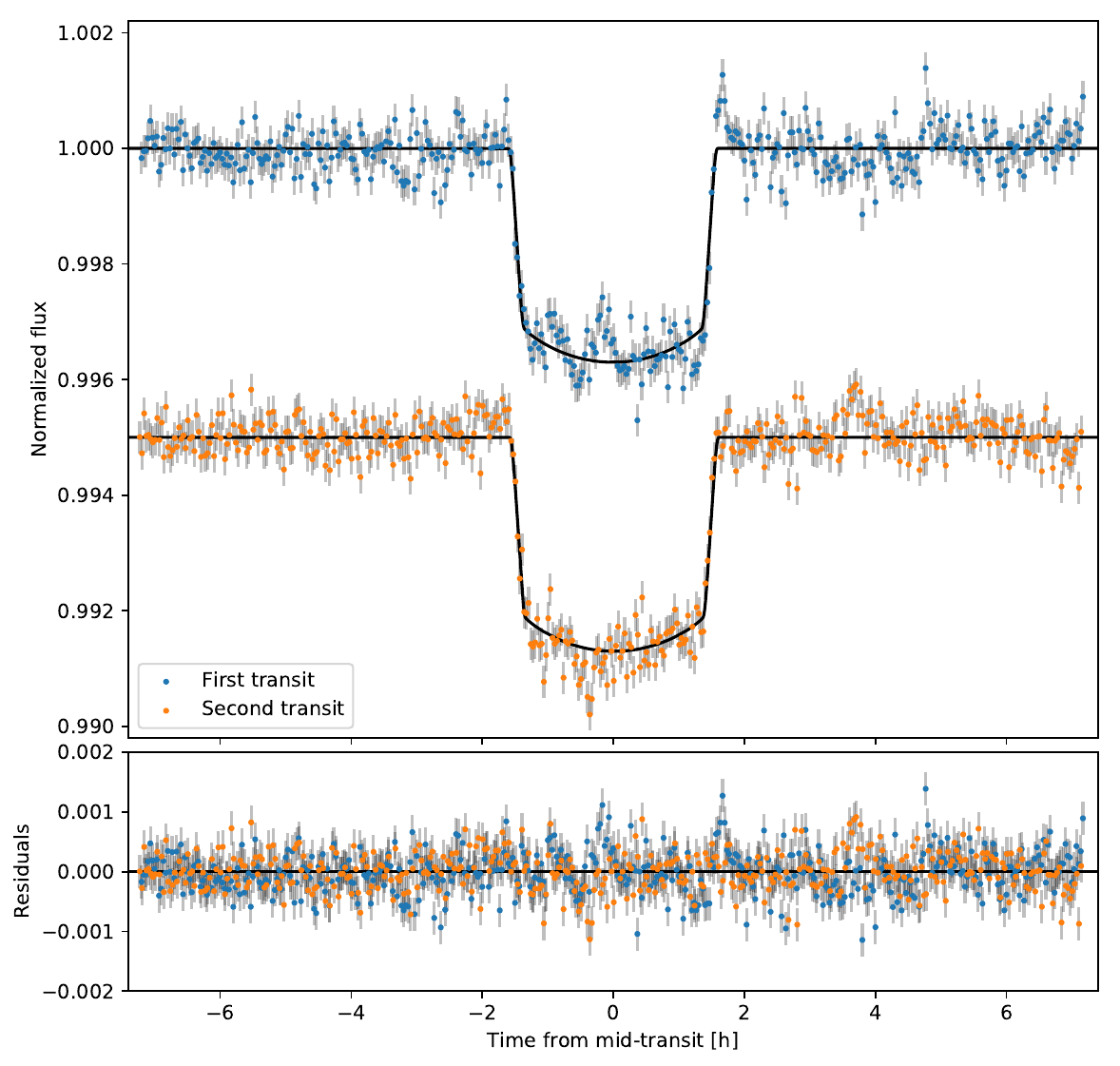}
  \caption{Upper panel: TESS light curves of the two transits of DS\,Tuc~A~b and the adopted fit (black line). Lower panel: residuals of the light curves after the subtraction of the model. \label{fig:transit_fit}}
\end{figure}

To improve the precision of both the mid-transit ephemeris and the orbital period, we repeated the same modelling by including the recovered third transit in the fit (as well as a third second-order polynomial for the detrending). The results are fully consistent with the previous model and returned the same uncertainties. Since the inclusion of the third transit doesn't improve the parameter determination, and the light curve is probably affected by instrumental systematics, we considered this model only for the estimation of T$_0$ and of the orbital period.
We reported in Table \ref{t:pl_param} the values obtained when assuming a circular orbit for the planet and no priors on the LD coefficients, while the resulting model is overimposed on the TESS light curves in Fig. \ref{fig:transit_fit} (black line). The value of the first coefficient of the two second order polynomials used to correct the raw transit light curves reflects the inclusion of the dilution factor.
The amount of jitter of the first transit is higher than the second one, suggesting that the variation of the amplitude of the light curve within the time-span of the TESS observation could be due to a variation of the stellar activity (cf. Fig. \ref{fig:cotrend}). 
The resulting planetary radius is $5.6 \pm 0.2$ R$_{\oplus}$, after taking into account the uncertainty on the stellar radius. This value decreases by $\sim 20\%$ when we neglect the dilution effect.
As in the case of the LD coefficients, above mentioned, magnetic activity can also affect the evaluation of other parameters of the star. For instance, the stellar radius can be inflated up to 10\%, as pointed out by \citet{2007A&A...472L..17C} and \citet{2007ApJ...660..732L}, with a possible impact on the planetary radius estimation. In the worst case, we could expect an increasing of the planet radius of about 15-20\%. 

\begin{table}[htbp]
   \caption[]{Parameters for DS\,Tuc~A~b.}
     \label{t:pl_param}
     \footnotesize
     \begin{center}
       \begin{tabular}{lll}
         \hline
         \noalign{\smallskip}
         Parameter   & Prior & Value\\
         \noalign{\smallskip}
         \hline
         \noalign{\smallskip}
\textit{Fitted:} & & \\
         \noalign{\smallskip}
Period [d]$^*$ & $\mathcal{U}$ (8.075,8.275)  & $8.1387 \pm 0.0005 $  \\
         \noalign{\smallskip}
T$_0$ [BJD-2450000]$^*$ & $\mathcal{U}$ (8332.2,8332.4)  & $8332.31013 \pm 0.00037$ \\
         \noalign{\smallskip}
R [R$_{\rm p}$/R$_{\star}$] & $\mathcal{U}$ ($1 \cdot 10^{-5}$,$5 \cdot 10^{-5}$) & $0.05972^{+0.00063}_{-0.00066}$ \\
         \noalign{\smallskip}
b & $\mathcal{U}$ (0.0,1.0)  &  $0.40^{+0.07}_{-0.08} $\\
         \noalign{\smallskip}
$\rho_{\star}$ [$\rho_{\star}$/$\rho_{\odot}$] & $\mathcal{N}$ (1.446,0.144)  &  $1.486^{+0.147}_{-0.137} $\\
         \noalign{\smallskip}
dilution factor & $\mathcal{N}$ (0.460,0.014) &  $0.459 \pm 0.014$\\
         \noalign{\smallskip}
jitter t$_1$ &  $\mathcal{U}$ ($3 \cdot 10^{-6}$, 0.028) &  $2.8\cdot 10^{-4} \pm 2\cdot 10^{-5}$ \\
         \noalign{\smallskip}
jitter t$_2$ &  $\mathcal{U}$ ($3 \cdot 10^{-6}$, 0.028) &  $1.7\cdot 10^{-4} \pm 2\cdot 10^{-5}$\\
         \noalign{\smallskip}
c$_0$ t$_1 $ &  $\mathcal{U}$ (-2.0,2.0) & $0.686 \pm 0.006$ \\
c$_0$ t$_2 $ &  $\mathcal{U}$ (-2.0,2.0) & $0.689 \pm 0.006$ \\
         \noalign{\smallskip}
c$_1$ t$_1 $ & $\mathcal{U}$ (-1.0,1.0) & $-0.0319 \pm 0.0003$ \\
c$_1$ t$_2 $ & $\mathcal{U}$ (-1.0,1.0)  & $0.0051 \pm 0.0001$ \\
         \noalign{\smallskip}
c$_2$ t$_1 $ &  $\mathcal{U}$ (-1.0,1.0) & $-0.0249 \pm 0.0006$ \\
c$_2$ t$_2 $ &  $\mathcal{U}$ (-1.0,1.0) & $-0.0123 \pm 0.0005$ \\
         \noalign{\medskip}
\textit{Derived:} & & \\
         \noalign{\smallskip}
a/R$_{\star}$ & --  &  $19.42 \pm 0.62$ \\
         \noalign{\smallskip}
LD$_{\rm c1}$ & -- &  $0.18^{+0.12}_{-0.11}   $\\
         \noalign{\smallskip}
LD$_{\rm c2}$ &  -- & $0.06^{+0.18}_{-0.15}$ \\
         \noalign{\smallskip}
inclination [deg] & --  &  $88.83^{+0.28}_{-0.24}$  \\
         \noalign{\smallskip}
Radius [R$_{\rm J}$]  &  -- & $0.50 \pm 0.02$ \\
         \noalign{\smallskip}
Radius [R$_{\rm \oplus}$] & -- & $5.63^{+0.22}_{-0.21} $ \\
         \noalign{\smallskip}
         \hline
      \end{tabular}
      \end{center}
$^*$ obtained with 3 transits fit
\end{table}

\subsection{Alternative modelling of three transit signals} \label{sec:fit2}
\begin{figure}[htbp]
  \includegraphics[width=\columnwidth]{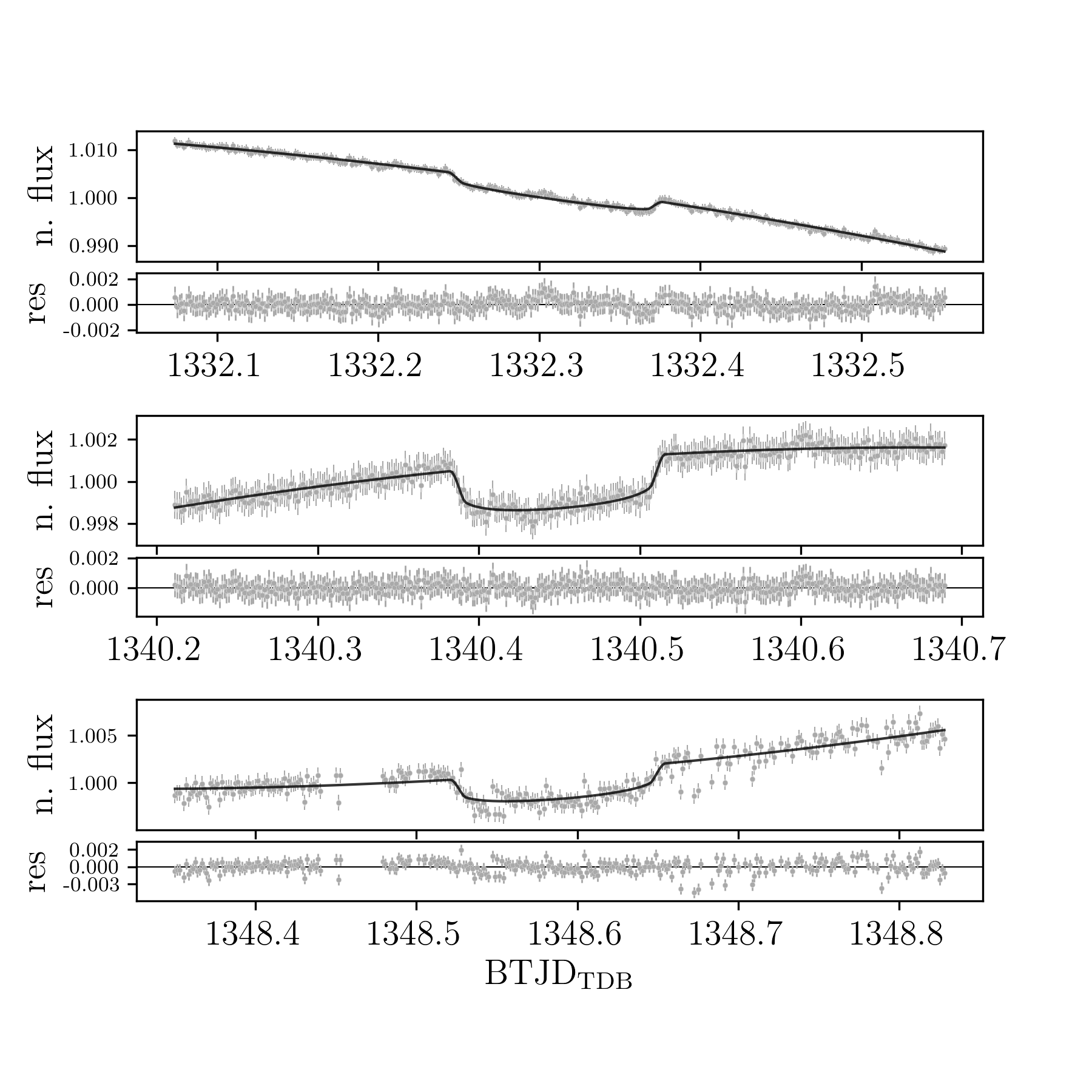}
  \caption{TESS undetrended light curves of the three transit events of DS\,Tuc~A~b, with the corresponding fit. \label{fig:3transits}}
\end{figure}
In our alternative approach for transit modelling, we flattened the extracted light curve with different approaches and window size:
running median, running polynomial of different order,
and a cubic spline on knots.
We applied the flattener algorithm after splitting the light curve at the TESS downlink time.
We determined that the best flattened light curve, having the smallest 68.27-th percentile scatter around the median value,
is a running polynomial of 3rd order with a window size of 0.55 days.
with respect to the median of the normalised flux.
This scatter metric has been used to remove outliers with an asymmetric sigma clipping:
$5\sigma$ above and $10\sigma$ below the median of the flux.
Also, the deviation measures the photometric scatter of the light curve and
we used it as the error on the light curve.
We searched for the transit signal on the flattened light curve with {\tt TransitLeastSquares}
(TLS, \citealt{2019A&A...623A..39H}).
We found a transit-like signal with a period of $8.13788 \pm 0.03704$~d,
a depth of 1830~ppm, a total duration ($T_{14}$) of about 172~min,
and a signal detection efficiency (SDE, \citealt{2000ApJ...542..257A}) of 17.58,
that is higher than the thresholds reported in \cite{2019A&A...623A..39H}.
The TLS found three transits in the light curve, and we selected a portion of  $\pm2 \times T_{14}$ around each transit time ($T_0$)
of the extracted light curve (not flattened).
We simultaneously modelled the three transits with the \texttt{batman-package} \citep{Kreidberg2015} and
analysed the posterior distribution with \texttt{emcee} \citep{Foreman-Mackey2013}.
We fit as common parameters across the three transits a dilution factor parameter (ratio between the flux of the secondary and the primary star, $F_B/F_A$),
the stellar density in gr cm$^{-3}$,
the logarithm in base two of the period of the planet ($\log_2P$),
the ratio of the radii ($k = R_\mathrm{p}/R_\star$),
the impact parameter ($b$),
the parameters $q_1$ and $q_2$ introduced in \citet{Kipping2013} for a quadratic limb darkening law,
the logarithm in base two of a jitter parameter ($\log_2\sigma_j$),
the reference transit time ($T_{0,\mathrm{ref}}$).
For each transit light curve,
we fitted three coefficients for a quadratic detrending ($c_0$, $c_1$, $c_2$),
and a transit time ($T_0$).
All the fitted and physical parameters have been bounded to reasonable values.
In particular the period was bounded within $8.13788 \pm 10 \times 0.03704$~d,
all the $T_0$ were limited within TLS guess $\pm T_{14}$~BTJD, and
we used the central transit times to bound the $T_{0,\mathrm{ref}} \pm T_{14}$.
We assumed a circular orbit for the planet.
We used the stellar parameters from Section \ref{sec:res} to determine an asymmetric Gaussian prior on the stellar density ($\mathcal{N}(2.00,[-0.21,+0.24])$).
The prior on the dilution factor has been computed as in Sec. \ref{sec:fit}.
We ran \texttt{emcee} with 60 chains (or walkers) for 50000 steps,
we discarded the first 20000 steps as burn-in
after checking the convergence of the chains (visual inspection and Gelman-Rubin $\hat{R} <= 1.02$).
We obtained the parameter posteriors after applying a pessimistic thinning factor of 100.
We computed the confidence intervals (CI) at the 16-th and 84-th percentile
and the high density intervals (HDI, corresponding at the 16-th and 84-th CI percentile).
The parameters, determined as the maximum log-likelihood estimator (MLE) within the HDI and the median of the posterior distributions, are reported in Tab \ref{tab:3fit} and are consistent within $1\sigma$
with the analysis presented in Sect. \ref{sec:fit}. Even in this case, a model with no priors on LD coefficients is preferred, according to the BIC (Bayesian Information Criterion).
The three transits with the median best-fit model are shown in Fig. \ref{fig:3transits}.

\section{Planet validation} \label{sec:val}

\subsection{Radial velocity and other spectroscopic time series} \label{sec:rv}
Five spectra of DS\,Tuc~A, spanning 100 days, were obtained in 2005 with HARPS (at the 3.6m telescope at ESO - La Silla) and the data, reduced with the standard pipeline, are available from the ESO Archive. Three additional spectra have been recently collected (Nov 2018 and Jun 2019) and included in our dataset. However, the HARPS standard reduction, performing the Cross-Correlation Function (CCF) of the spectrum with a digital mask representing the typical spectral features of a G2 star \citep{2002A&A...388..632P}, is not suitable for a fast rotator as DS\,Tuc~A, so we reprocessed the spectra with the offline version of the HARPS data reduction software (DRS, C. Lovis private communication), where the width of the CCF can be adapted according to the rotational broadening. In this case, we used 240 km s$^{-1}$. We also employed the TERRA \citep{2012ApJS..200...15A} and SERVAL \citep{2018A&A...609A..12Z} pipelines to obtain alternative RV measurements for comparison. 

When joining the two datasets, separated by 13 years, we must take into account an RV offset produced by a change in the HARPS set-up occurred in 2015, which is in average $\sim 16$ m s$^{-1}$ for G-type slowly-rotating stars, as reported by \citet{locurto2015}. In particular, these authors found a dependence of the RV shift on the width of the spectral lines, that in the case of DS\,Tuc~A are particularly broadened. However, a comparison between old and new RV measurements for different type of stars, including young objects and fast rotators, shows that the average offset is compatible within errors with respect to the result in \citet{locurto2015}. In our case, we found an offset between the two series of about 32 $\pm$ 77 m s$^{-1}$, significantly lower than the scatter of the data.

\begin{figure}[htbp]
\includegraphics[width=0.7\columnwidth,angle=90]{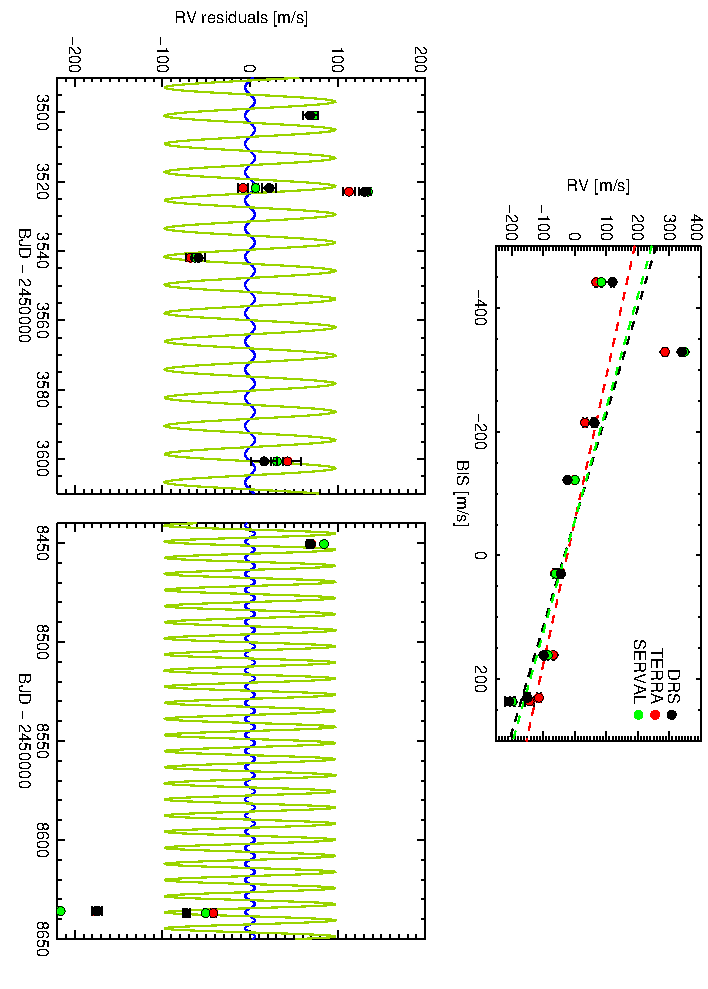}
\caption{\label{fig:rv} 
Upper panel: Correlation between HARPS bisector span and RVs obtained from the three different pipelines.
Lower panels: Time series of the RV residuals after removing the correlation from HARPS RVs (same color code for the pipelines). The solid lines represent the RV signal in case of a planet with the same orbital elements as DS\,Tuc~A~b and a semi-amplitude corresponding to the mass upper limit estimated from the residuals (green line) and from empirical relation (blue line).}
\end{figure}

The RV dispersion of the time series is $\sim 174$ m s$^{-1}$ for both HARPS offline DRS, $\sim 136$ m s$^{-1}$ for TERRA (the lower dispersion is due to the fact that the algorithm used by TERRA better models the distorted line profile of both M dwarfs and active stars, and thus is particularly suitable to obtain RVs for this kind of objects, as already shown by \citealt{2017A&A...598A..26P}) and 172 m s$^{-1}$ for SERVAL, with corresponding median RV error of 7.6 m s$^{-1}$, 6.2 m s$^{-1}$ and 3.1 m s$^{-1}$.
Finally, we measured several activity indicators: the bisector span and the $\log R'_{HK}$, directly provided by the HARPS DRS, H$\alpha$ (as in \citealt{2011A&A...534A..30G}), the chromatic index (as derived by SERVAL), and $\Delta V$, $V_{{\rm asy,mod}}$ and an alternative evaluation of the bisector through the procedure presented in \citet{2018A&A...616A.155L}.
All of these indicators show a clear anti-correlation with respect to the RVs obtained with all the pipelines, as expected for a young and active star, for which the rotation dominates the RV signal (see the upper panel of Fig. \ref{fig:rv}). An outlier is present in the time series, not related to the processing method since all the pipelines report the same anomalous value. On the other hand, there is no indication of flare or signal of peculiar activity in that spectrum, or problems in data collection (such as low signal-to-noise ratio, high airmass or lunar contamination). For this reason, we included this datum in our analysis.
The Spearman correlation coefficient, between those RVs and the bisector is $\sim -0.97$ for the three datasets, with a significance of $\sim 2 \cdot 10^{-4}$, evaluated through the IDL routine {\tt SAFE\_CORRELATE}.
When we subtract this correlation from the RVs (e.g., HARPS DRS dataset), the dispersion of the residuals decreases from $\sim 200$ to $\sim 100$ m s$^{-1}$, from which we obtain a first approximation of the upper limit on the planet real mass (since we know the inclination of the orbit $i = 88.8^{\rm o}$), that would be around 1.3 M$_{\rm Jup}$. 
It is reasonable to expect that those residuals are still affected by an additional contribution of the stellar activity, and a significantly lower value of the planetary mass is more likely.
Indeed, according to the empirical mass-radius relations provided by \citet{2017A&A...604A..83B}, we should expect a planetary mass of $22.9 \pm 2.5$ M$_{\oplus}$. However, we must take into account that those relations are deduced from a sample of well-characterised exoplanets, generally hosted by mature stars, that may not be applicable to young objects still experiencing their contraction phase.
By using the orbital elements from Table \ref{t:pl_param} and considering an RV semi-amplitude compatible with the estimated mass upper limit from our data and the mass from the empirical relation, we obtain RV models to compare with the residuals. Lower panels of Fig. \ref{fig:rv} shows that the RV residuals are compatible with those planetary masses (green and blue line, respectively). However, we stress that a full analysis of the activity is not fully reliable with the available data, due to their sparse sampling. 
\begin{table}[htbp]
\centering
\caption{Radial velocities from DS\,Tuc~A HARPS spectra with uncertainties as obtained from the HARPS offline DRS and the corresponding measure of the bisector span with uncertainties as derived from the procedure described in \citet{2018A&A...616A.155L}.}
\small
\begin{tabular}{lcccc}
\toprule
 BJD$_{\rm TDB}$ - 2450000    & RV   &    err$_{\rm RV}$     &   BIS  &  err$_{\rm BIS}$  \\
                              & [km s$^{-1}$]    &  [km s$^{-1}$]   &  [km s$^{-1}$]  &  [km s$^{-1}$]   \\
\midrule
3500.876233  & 7.7420  & 0.0087  & 0.2301  & 0.0217  \\
3521.828165  & 7.8486  & 0.0083  & 0.0296  & 0.0439  \\
3522.888133  & 8.2321  & 0.0063  & -0.3288 & 0.0249  \\
3541.927465  & 7.9548  & 0.0076  & -0.2145 & 0.0420  \\
3600.704290  & 7.6846  & 0.0146  & 0.2361  & 0.0364  \\
8450.526567  & 7.8270  & 0.0033  & 0.1610  & 0.0144  \\
8635.891307  & 8.0455  & 0.0052  & -0.4416 & 0.0880  \\
8636.897996  & 7.9027  & 0.0043  & -0.1219 & 0.0169  \\
\bottomrule
\end{tabular}
\label{tab:rv}
\end{table}

\begin{figure}[htbp]
\includegraphics[width=\columnwidth,angle=0]{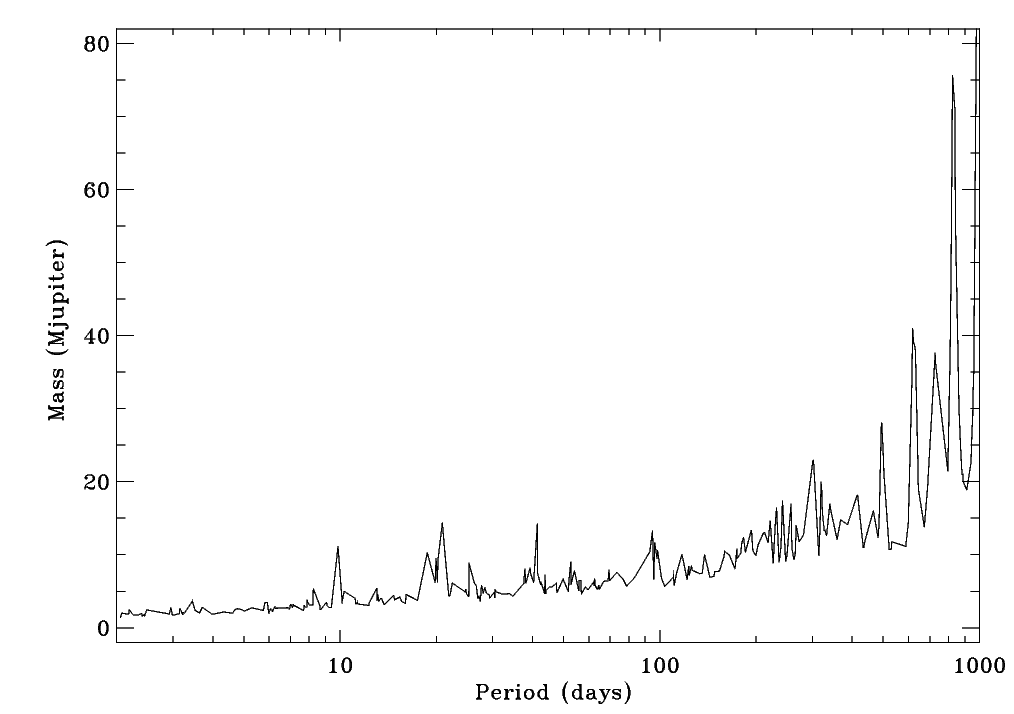}
\caption{\label{fig:detlim} 
Detection limits of the RV time series of DS\,Tuc~A with a confidence level of 95\%.}
\end{figure}

As a further check, by using the same RV residuals we
evaluated the corresponding period-mass detection limits,
based on the comparison of the variance of the RV time series with and without
a variety of planet masses and periods added to the data, using the F
test as in \citet{2003A&A...405..207D}.
Fig. \ref{fig:detlim} shows that we can rule out the existence of
planetary companions with m$\sin i$ larger than 2-3 M$_{\rm Jup}$ with an orbital period shorter than 10 days, while the detection threshold is
less sensible (M $\sim$ 7 M$_{\rm Jup}$) at about 100 days.

The observed RV dispersion of DS\,Tuc~A is within the range of dispersion for stars
with similar age and spectral type, as derived from HARPS spectra of several objects available in the ESO archive, further supporting
the magnetic activity as the dominant source of variability (e.g., \citealt{kraus2014}).
The CCF and line profile indicators also show the typical
alterations due to the presence of active regions, without any indication of double-lined spectroscopic binary.

High-resolution HARPS spectra are available only for the primary star.
However, RV monitoring the secondary is available from \citet{torres2006},
resulting in a low dispersion of 0.1 km/s (for six spectra obtained with FEROS), 
ruling out the presence of close stellar companions also for the secondary.
More in general, the similar absolute RVs reported by
several authors for both components (see Table \ref{t:star_param})
argue against any RV variation exceeding 1 km/s over a baseline of decades.

\subsection{Constraints from direct imaging} \label{sec:imag}
\begin{figure}[htbp]
\centering
\includegraphics[width=\columnwidth]{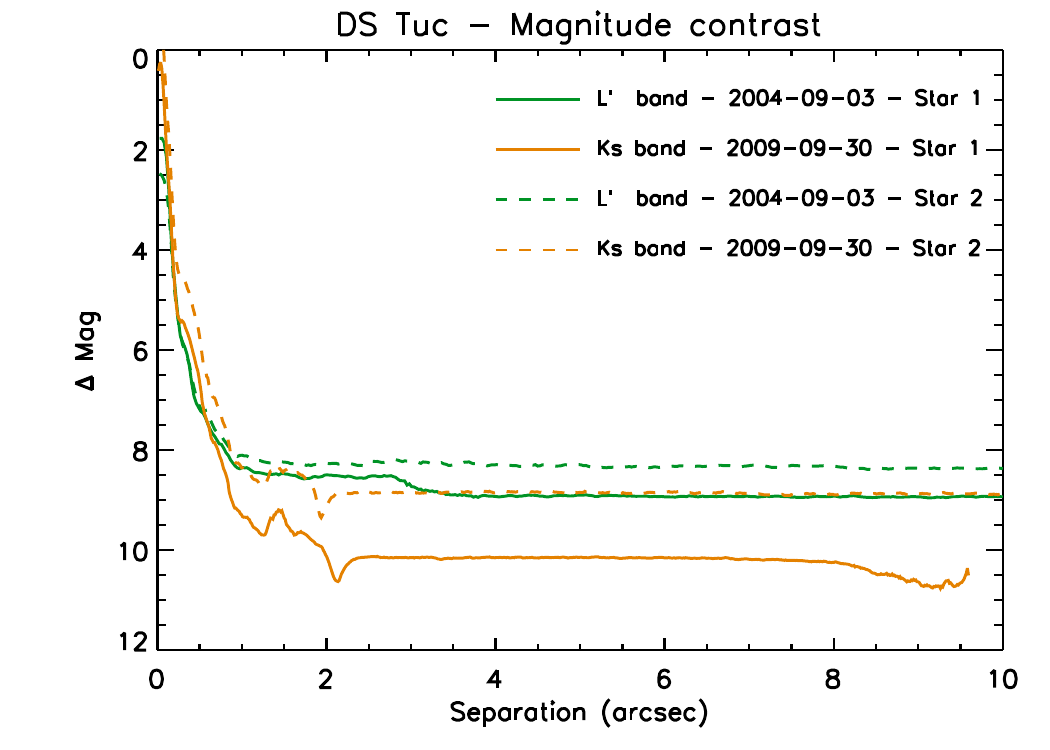}
\caption{\label{fig:magcontrast} Contrast plot in $\Delta$Mag 
obtained around both stars of the binary both for the 2004 and 2009 observation.}
\end{figure}

We exploited the NaCo (Nasmyth Adaptive Optics System Near-Infrared Imager and Spectrograph, mounted at the Very Large Telescope, ESO - Paranal Observatory, Chile) datasets in L' spectral band  taken on 2004, Sept 3rd and published in \cite{kasper2007} and in Ks band taken on 2009 Sept 30th and published in \cite{vogt2015}.
Both datasets were reduced following the methods devised in the
original papers and were then registered in such a way that the 
primary star was at the center of the image. To improve the
contrast in the regions around the two stars we subtracted each
other the PSF of the two stars following the method presented
in \citet{desidera11} and in \citet{carolo14}. The contrast limit
was then calculated using the same method described in \citet{Mesa15}. We display the contrast plots for both epochs and for both stars in Figure~\ref{fig:magcontrast}. 

\begin{figure}[htbp]
\includegraphics[width=\columnwidth]{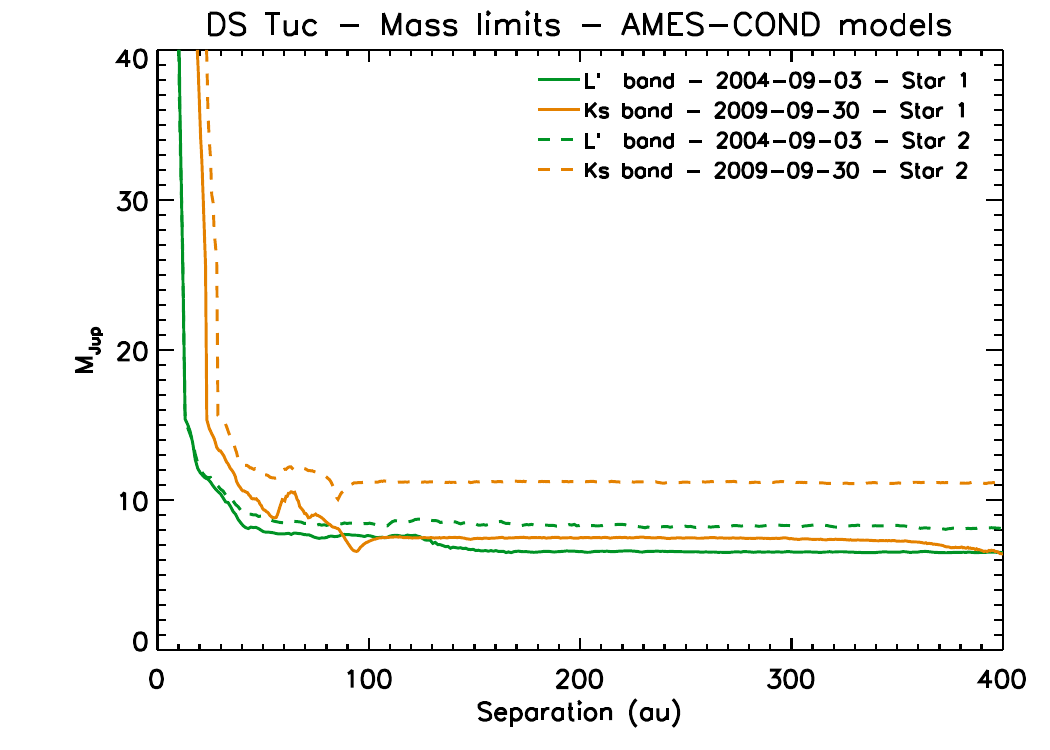}
\caption{\label{fig:masslimit} Mass detection limits expressed in
$M_{Jup}$ as a function of the separation from the stars expressed
in au for both observing epochs and for both the stars of the binary.}
\end{figure}
We used these contrast values to define mass detection limits around both
stars of the binary, assuming age and distance given in Table~\ref{t:star_param} and exploiting the AMES-COND evolutionary models \citep{allard03}. The final results of this procedure are
displayed in Figure~\ref{fig:masslimit} for both the observing epochs
and for both the stars of the binary. For what concerns the primary
star, we are able to exclude the presence of objects with mass larger than $\sim$7-8~$M_{\rm Jup}$ at separations larger than 40~au.
The mass limits around the secondary are slightly larger.

The $\sim$ 5 yr time baseline between the two NaCo observations coupled with the proper motion and parallax of the star implies a relative shift of 524 mas for
a stationary background object with respect to DS Tuc.
Therefore, from the detection limits in Fig \ref{fig:magcontrast},
a stationary background source responsible for the transit feature
($\Delta mag$ smaller than about 5.5 mag assuming 0.75 mag eclipse depth)
would have been detected in at least one of the two epochs 
and is then ruled out by the observations.
The only remaining false alarm sources are represented by bright stellar companions at projected separation 
smaller than $\sim$10 au but large physical separations, considering the detection limits from RVs and imaging. This configuration is extremely unlikely from the geometrical point of view and is made negligible by the presence of DS Tuc B at 240 au.
The imaging observation reported here and the analysis of sources
within TESS PSF from Gaia DR2 also do not allow us to identify reliable
candidates for being responsible of the false alarm candidate reported
in the original TESS alert. Most likely the corresponding photometric signal
is not of astrophysical origin.

Coupling the results of the RV monitoring and the NaCo observations, the presence of an additional companion responsible for the photometric dimming observed with TESS is extremely unlikely. As resulting from Gaia DR2, there are no additional sources brighter than G=18.3 (10 mag fainter than the primary) within 1 arcmin.

\subsection{False positive probability} \label{sec:fpp}
\begin{figure*}[htbp]
\centering
  \includegraphics[width=1.5\columnwidth]{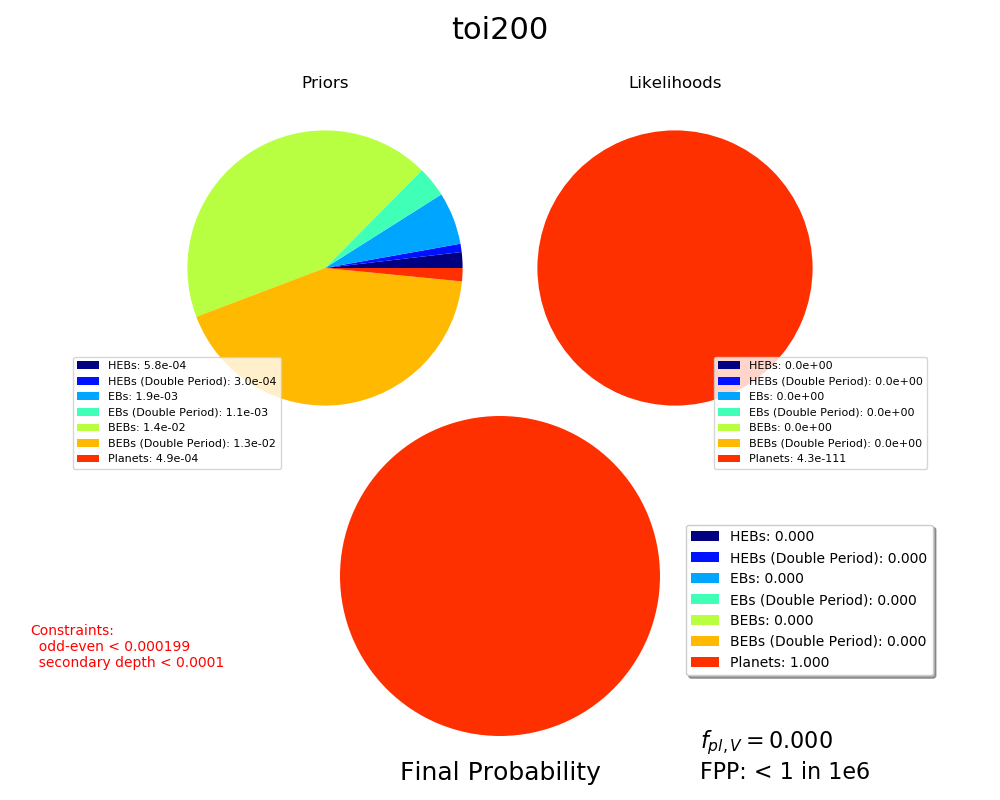}
  \caption{Summary of the FPP evaluation obtained with {\tt vespa}. Acronyms of the false positive scenarios are defined in Sect.\ref{sec:fpp} \label{fig:fpp}}
\end{figure*}

As a final step, we run the free and open source validation tool {\tt vespa} \citep{2012ApJ...761....6M,2015ascl.soft03011M} to evaluate the false positive probability (FPP) of DS\,Tuc~A~b in a Bayesian framework. {\tt vespa} has been widely used (e.g., \citealt{2015ApJ...809...25M, 2016ApJS..226....7C,Malavolta2018,2018AJ....155..136M,2019MNRAS.487.1865W}) to estimate the probability that a given signal is actually produced by a real planet and not the result of different astrophysical configurations. As reported in \citet{2012ApJ...761....6M}, the false positive scenarios explored by {\tt vespa} are: \textit{i)} a non-associated back/foreground eclipsing binary blended inside the photometric aperture of the target (BEB), \textit{ii)} a hierarchical triple system where two components eclipse (HEB), \textit{iii)} an eclipsing binary (EB), \textit{iv)} a non-associated blended back/foreground star hosting a transiting planet (Bpl). 
For each configuration, the code simulates a representative stellar population, that in our case is constrained by the information provided in Table \ref{t:star_param} and \ref{t:pl_param} (coordinates, optical and near-infrared photometry, orbital period and the ratio between planet and star radius) and the observed TESS light curve and NaCo contrast curve. Different false positive scenarios are considered for those populations, and are used to define the prior likelihood that a specific configuration actually exists (i.e., that is consistent with the input data) and the likelihood of transit for those configurations. The comparison with the full set of scenarios allows to evaluate the FPP: if the FPP is significantly lower than 0.01, the planet is validated.
We run {\tt vespa} with the available data of DS\,Tuc~A~b, obtaining an FPP lower than $10^{-6}$, resulting in a full validation of the transiting planet scenario. Curiously, a large fraction of BEB configurations are in principle permitted by the data (upper left pie chart in Fig. \ref{fig:fpp}), but the fitting of the input transit signal clearly favours the planet hypothesis (upper right pie chart in Fig. \ref{fig:fpp}).

\section{Discussion and Conclusions}
\label{sec:conc}
In this paper, we have ruled out blending scenarios and then validated the planet candidate found with TESS around the primary component of the young binary system DS\,Tuc, confirmed member of the Tuc-Hor association. We estimated the transit parameters, in particular, the planet radius, taking into account the dilution effect due to the stellar companion of the host star. 
The comparison with theoretical models, accounting for the young age of the hosts, by \citet{2003A&A...402..701B} and \citet{linder2018} provides a mass below 20 $M_{\oplus}$, compatible with the empirical estimation of $\sim 22$ M$_{\oplus}$, suggesting that  DS\,Tuc~A~b could be a super-Earth or a Neptune-like planet at most. However, we stress that these models are not optimised for heavily irradiated planets around young stars, i.e., they don't consider the possible inflation of the planetary radii. 

The relatively low mass expected for the planet makes the detection of the RV signature quite challenging considering the amplitude of the activity jitter, as shown in Sect. \ref{sec:rv}. According to our simulations, an intensive monitoring (e.g., $\sim 60$ RV epochs spanning three months), coupled with appropriate modeling of the activity would allow the recovery of a signal with RV semi-amplitude down to $\sim 6.5$ m s$^{-1}$ (corresponding to 20 M$_{\oplus}$) with an accuracy of 1-$\sigma$ and a detection significance of 2-$\sigma$, in case of an activity signal of the order of 100 m s$^{-1}$.
More promising is the perspective for the detection of the Rossiter-McLaughlin
effect (RM), as its amplitude depends on planetary radius and not on planetary mass and increases with the projected rotational velocity of the star. Furthermore, the activity noise is expected to be much smaller on the timescales of a single transit with respect to the few weeks/months needed for the orbit monitoring. 
Assuming the values in Table \ref{t:star_param} and \ref{t:pl_param}, the RM signature should be around $\sim 60$ m s$^{-1}$ (Eq. 40 in \citealt{2010arXiv1001.2010W}).
Considering the moderately long period of the planet and the young age of the system, the spin-orbit angle 
is expected to be close to the original one, with little alterations by tidal effects. 
The measurement of the RM effect would then be extremely insightful of the migration history of the planetary system.
Furthermore, the binarity is an additional reason for specific interest for the RM determination. Finally, due to the large value of the v$\sin i$, this target can be studied with the tomographic technique.

The expanded structure of the planet favours the detection of strong atmospheric features. The scale height, obtained according to the equilibrium temperature ($\sim 900$K) is 1200 km, so the expected signal of the transmission spectrum feature is about $2 \cdot 10^{-4}$.
According to our evaluations, DS\,Tuc~A~b is the first confirmed transiting planets around stars with age $\sim 40$ Myr with the potentiality of a full characterization.

\begin{acknowledgements}
Authors thank the anonymous referee for her/his interesting comments and suggestions that improved the quality and robustness of the paper. We acknowledge the support by INAF/Frontiera through the "Progetti Premiali" funding scheme of the Italian Ministry of Education, University, and Research. Authors acknowledge Dr. A. Santerne, Dr. A.~F. Lanza and Dr. F. Borsa for their kind and useful suggestions. LM acknowledges support from  PLATO ASI-INAF agreement n.2015-019-R.1-2018. LBo acknowledges the funding support from Italian Space Agency (ASI) regulated by Accordo ASI-INAF n. 2013-016-R.0 del 9 luglio 2013 e integrazione del 9 luglio 2015. 
ME acknowledges the support of the DFG priority programSPP 1992 "Exploring  the  Diversity  of  Extrasolar  Planets" (HA 3279/12-1). We acknowledge the use of public TESS Alert data from pipelines at the TESS Science Office and at the TESS Science Processing Operations Center.
Funding for the TESS mission is provided by NASA’s Science Mission directorate.
During the final stages of preparation of this paper, we were learned of an independent analysis done by another team \citep{2019arXiv190610703N}.
\end{acknowledgements}


%

\appendix
\section{}
\label{a:table}

\begin{table}[htbp]
\centering
\caption{Parameters of the alternative 3-transits fit presented in Sec. \ref{sec:fit2}. Uncertainties are defined as 1-sigma of the high density intervals}
\begin{tabular}{ll}
\toprule
\noalign{\smallskip}
Parameter  & Value   \\
\noalign{\smallskip}
\midrule
         \noalign{\smallskip}
\textit{Fitted}:  &    \\ 
\noalign{\smallskip}
dilution factor  & 0.4602$_{-0.0179}^{+0.0116}$   \\
\noalign{\smallskip}
$\rho_{\star}$ [gr cm$^{-3}$] & 1.9846$_{-0.2458}^{+0.2612}$ \\
\noalign{\smallskip}
$\log_2P$   &  3.0248$\pm 0.0001$ \\
\noalign{\smallskip}
R$_{\rm p}$/R$_{\star}$  &  0.0606$_{-0.0011}^{+0.0009}$\\
\noalign{\smallskip}
Impact parameter $b$                &    0.4356$_{-0.0942}^{+0.0760}$\\
\noalign{\smallskip}
LD$_{\rm q1}$      &   0.1414$_{-0.0889}^{+0.1203}$\\
\noalign{\smallskip}
LD$_{\rm q2}$      &  0.1841$_{-0.1841}^{+0.1053}$\\
\noalign{\smallskip}
$\log_2 \sigma_{\rm j}$   &  -11.2057$_{-0.0730}^{+0.0759}$\\
\noalign{\smallskip}
T$_0$ [BJD-2,450,000] & 8340.44905$_{-0.00039}^{+0.00042}$\\
\noalign{\smallskip}
TT$_0$ [BJD-2,450,000]  &  8332.31019$_{-0.00048}^{+0.00072}$ \\
\noalign{\smallskip}
TT$_1$ [BJD-2,450,000]  &  8340.44905$_{-0.00039}^{+0.00042}$ \\
\noalign{\smallskip}
TT$_2$ [BJD-2,450,000]  & 8348.58791$_{-0.00087}^{+0.00064}$ \\
\noalign{\smallskip}
c$_0$ transit 1    &    0.0112$\pm 0.0001$ \\
\noalign{\smallskip}
c$_1$ transit 1    &  -0.0283$\pm 0.0001$ \\
\noalign{\smallskip}
c$_2$ transit 1    &  -0.0392$\pm 0.0020$ \\
\noalign{\smallskip}
c$_0$ transit 2    &  -0.0016$\pm 0.0001$ \\
\noalign{\smallskip}
c$_1$ transit 2    &   0.0156$_{-0.0014}^{+0.0010}$ \\
\noalign{\smallskip}
c$_2$ transit 2    &  -0.0170$_{-0.0023}^{+0.0025}$ \\
\noalign{\smallskip}
c$_0$ transit 3    &  -0.0006$\pm 0.0001$ \\
\noalign{\smallskip}
c$_1$ transit 3    &   0.0014$\pm 0.0001$ \\
\noalign{\smallskip}
c$_2$ transit 3    &   0.0242$_{-0.0024}^{+0.0022}$ \\
\noalign{\smallskip}
\textit{Derived}:  &    \\ 
\noalign{\smallskip}
Period [days]     &  8.1389$_{-0.0004}^{+0.0006}$ \\
\noalign{\smallskip}
(R$_{\rm p}$/R$_{\star})_{\rm obs}$ &   0.0502$\pm 0.0008$ \\
\noalign{\smallskip}
Radius [R$_{\rm \oplus}$]    &  5.7961$_{-0.1061}^{+0.0885}$\\
\noalign{\smallskip}
inclination [deg]  &   88.6921$_{-0.3721}^{+0.2271}$\\
\noalign{\smallskip}
a/R$_{\star}$       &  19.0834$_{-0.8821}^{+0.7662}$ \\
\noalign{\smallskip}
T$_{14}$ [min]  & 189.1796$_{-1.9321}^{+1.5862}$\\
\noalign{\smallskip}
LD$_{c1}$ (linear)      & 0.1385$_{-0.1385}^{+0.0536}$\\
\noalign{\smallskip}
LD$_{c2}$ (quadratic)   & 0.2375$_{-0.2375}^{+0.1249}$ \\
\noalign{\smallskip}
$\sigma_{\rm jitter}$   & 0.00040$ \pm 0.00002$ \\
\noalign{\smallskip}
\bottomrule
\end{tabular}
\label{tab:3fit}
\end{table}

\end{document}